# Diffractive Magic Cube Network with Super-high Capacity Enabled by Mechanical Reconfiguration

Peijie Feng[1], Fubei Liu[2], Yuanfeng Liu[2], Mingzhe Chong[1], Zongkun Zhang[1], Qian Zhao[3], Jingbo Sun[2], Ji Zhou[2], Yunhua Tan[1]

[1]School of Electronics, Peking University, Beijing, 100871, China

[2]State Key Laboratory of New Ceramics and Fine Processing, School of Materials Science and Engineering, Tsinghua University, Beijing 100084, China

[3]State Key Laboratory of Tribology in Advanced Equipment, Department of Mechanical Engineering, Tsinghua University, Beijing, China

These authors contribute equally: Peijie Feng & Fubei Liu

Correspondence: Yunhua Tan Email: tanggeric@pku.edu.cn & Ji Zhou Email: zhouji@tsinghua.edu.cn& Jingbo Sun Email: jingbosun@mail.tsinghua.edu.cn

Peijie Feng Email: pjf@stu.pku.edu.cn

Fubei Liu Email: lfb23@mails.tsinghua.edu.cn

Yuanfeng Liu Email: lyf_student@126.com

Mingzhe Chong Email: cmz1999@stu.pku.edu.cn

Zongkun Zhang Email: zhangzongkun1998@163.com

Qian Zhao Email: zhaoqian@tsinghua.edu.cn

**Abstract**

Free-space wavefront manipulation devices have emerged as powerful platforms for advanced optical information systems. In response to the challenges posed by the exponential growth of optical information, optical multiplexing and dynamic reconfigurable devices are being actively explored to the enhance system capacity. Among them, coarse-grained mechanically reconfigurable mechanism offers a cost-effective and low-complexity approach for capacity enhancement. However, the channel numbers achieved in current studies are insufficient for practical applications because of inadequate mechanical transformations and suboptimal optimization models. In this article, a diffractive magic cube network (DMCN) is proposed to advance the multiplexing capacity of mechanically reconfigurable system. We utilized the diffractive deep neural network ($D^2$NN) model to jointly optimize the subset of channels generated by the combination of three mechanical operations, permutation, translation, and rotation. The 144-channel holograms, 108-channel single/double focus, 60-channel single/multi-mode OAM beam generation were experimentally demonstrated using diffractive optical elements (DOEs). An equivalent connectivity law was formulated to improve model scalability. Our strategy not only provides a novel paradigm to improve system capacity to super-high level with low crosstalk, but also paves the way for new advancements in optical storage, computing, communication, and photolithography.

**Introduction**

Free-space wavefront manipulation devices with different modulation pixel scales, such as diffractive optical elements (DOEs), metasurfaces, and liquid crystals (LCs), have provided powerful platforms for the development of novel optical information systems, including holography[1-5], flat lenses[6-8], orbital angular momentum (OAM) beam generation[9-12], optical computing[13-16] etc. The information capacity of these systems needs to be continually enhanced to accommodate the ever-growing demands of massive information processing. In recent years, extensive studies on optical multiplexing strategies have been conducted to enhance system capacity. Early efforts successfully develop independent multiplexing channels leveraging wavelength[2,17], polarization[18-21], OAM[22], and transmission direction[23] of light but these approaches provide limited channels (i.e., less than 4) due to the constraints on channel independence. Fortunately, Xiong et al. achieved 11 non-orthogonal polarization multiplexing channels by introducing correlated noise into the least squares algorithm[24]. This breakthrough overcomes the design limitations of independent channels and demonstrates the possibility of multiplexing using correlated channels. Inspired by this work, an increasing number of studies have begun to explore the non-orthogonal polarization channels for multifunctional multiplexing[25-27]. However, the correlation between multiple channels increases crosstalk, making the device parameter design more challenging. Diffractive deep neural network ($D^2$NN), a data-driven deep learning optical network is frequently used in these researches to address the obstacle. The state-of-the-art work reported by Wang et al. utilized $D^2$NN to jointly optimize 55 non-orthogonal polarization channels obtained by extending the dimension of the Jones matrix, achieving ultrahigh-capacity holography[27].

Dynamic wavefront manipulation devices with reconfigurable or reprogrammable features are also widely adopted to enhance the system capacity. Common strategies for dynamically controlling the optical response of materials involve introducing the external stimuli, such as voltage[28-31], light[32], and temperature[33]. These methods incorporate extra components, including PIN diodes, LCs, photodiodes, and $VO_2$, into the devices, thereby increasing system complexity and energy consumption. Instead, mechanically reconfigurable devices comprising multiple modulation layers offer a cost-effective and low-complexity solution for dynamic manipulation of light. These systems flexibly alter the relative spatial configuration of multiple modulation layers through operations such as rotation and insertion, thereby enabling coarse-grained reconstruction of the manipulation process[34–40]. Some studies designed compact cascaded devices using physics-inspired formulations or gradient descent algorithms[34–39]. With extremely short interlayer spacing, interlayer diffraction becomes negligible, and the multiple modulation layers effectively function as a single layer. This not only enhances inter-channel correlation but also significantly restricts the mechanical degrees of freedom (DOF). Leveraging the $D^2$NN framework, Fan et al. proposed the TDNN with increased interlayer spacing[40], ensuring sufficient diffractive connectivity between layers. The DOF in rotational reconfiguration were effectively enhanced through the design of circular rotation interpolation model. However, the increase in channel count unavoidably reduces the rotation step size, which in turn accelerates the growth of inter-channel correlation and interpolation-related numerical errors. And the circular rotation model confines the TDNN to PB phase implementations. These limitations hinder the scalability and portability of the TDNN. Overall, the experimentally achieved channel numbers in current studies remain limited (i.e., fewer than 20) due to inadequate mechanical transformations and ineffective optimization models, which are far from meeting the demands of practical applications. The increased channel number also reduces the signal-to-noise ratio (SNR), and a trade-off between them has yet to be well established. Therefore, the ultimate limit for this type of coarse-grained reconfigurable mechanisms remains to be explored.

In this article, we introduce a diffractive magic cube network (DMCN) consisting of three diffractive layers, elevating the multiplexing capacity of mechanically reconfigurable mechanisms to a super-high level. Our DMCN constructed thousands of non-independent potential channels through the combination of three coarse-grained mechanical operations: permutation, translation, and rotation. We carefully selected the subsets of these channels and employed the $D^2NN$ model to optimize the transmission phases of the DMCN, achieving 144-channel holograms (PCC ~ 0.99), 108-channel single/double focus (PSNR ~ 42.9 dB), and 60-channel single/multi-mode OAM beam (CC ~ 1). An equivalent connectivity law for the DMCN model was theoretically derived to improve its scalability. The interrelationships among the layer width, field of view (FOV) width, channel count and SNR were also analyzed in detail, revealing that an increased layer-to-FOV width ratio enhances channel capacity and SNR. We experimentally validated the numerical model using multi-layer DOEs fabricated via grayscale photolithography, and observed agreement between measurements and simulations. Since the DMCN was implemented by the simplest phase-modulated cells, it can easily be integrated with techniques such as polarization multiplexing and wavelength multiplexing to further enhance channel count. Our strategy provides a novel paradigm for enhancing system capacity, which involves constructing a large amount of non-independent channels across multiple dimensions and subsequently employing $D^2NN$ model for joint optimization. This work will provide new possibilities for the development of optical storage, display, encryption, photolithography[41] and optical communications[42,43].

## Results

### The super-high capacity and optimization of DMCN

The system schematic of our DMCN is shown in **Fig. 1**. We employed the forward propagation model of $D^2NN$ to describe the wave transfer procedure in the system. Specifically, the incident light is first modulated by the diffractive layer, and the resulting modulated wavefront then acts as a secondary wave source, radiating waves to the adjacent layer. This process is repeated three times, ultimately forming the target wavefield at the output FOV. A corresponding system transfer matrix $\boldsymbol{A_c}$ (see **Methods** for details) was also derived to help demonstrate the reconfigurable mechanism of DMCN. Its parameters can be altered by three types of coarse-grained mechanical operations, enabling the generation of potential channels. Permutation operations change the order of the layers (e.g., ABC→BAC), thereby altering the positions of the modulation coefficient matrices $\boldsymbol{T_A}, \boldsymbol{T_B}, \boldsymbol{T_C}$ in equation $(\text{e.g.}, \boldsymbol{T_1} = \boldsymbol{T_A}, \boldsymbol{T_2} = \boldsymbol{T_B}, \boldsymbol{T_3} = \boldsymbol{T_C} \rightarrow \boldsymbol{T_1} = \boldsymbol{T_B}, \boldsymbol{T_2} = \boldsymbol{T_A}, \boldsymbol{T_3} = \boldsymbol{T_C})$. Rotation operations applied to the second and third layers modify the rotation angles of the matrices of $\boldsymbol{T_2}$ and $\boldsymbol{T_3}$, denoted as $\Omega_1$ and $\Omega_2$. Translation operations adjust the layer distances $d_1, d_2, d_3$ through shifting the second and third layers. Notably, the total longitudinal length, $d_1 + d_2 + d_3$, is designed as a const, allowing the optical information stored in the different mechanical states of the DMCN to be excited and directed to the fixed output FOV under the illumination of incident plane wave $\boldsymbol{J}$.

**Fig. 2 b** illustrates the combination manners of aforementioned mechanical operations**,** which endows the DMCN with thousands of potential channels. The full permutations (i.e., $P_3^3 = 6$) and partial permutations (i.e., $P_3^2 + P_3^1 = 9$) of the three modulation layers generated 15 channels. The second and third layers of DMCN can be respectively rotated to four specific angles: 0°, 90°, 180°, and 270°, resulting in another 16 possible combinations. Lastly, the total longitudinal length of 60 cm was discretized into 60 intervals, each measuring 1 cm. The intervals allocated to $d_1$ and $d_2$ were equal to facilitate the experiment, which produced 29 possible configurations of interlayer distances for translation operations. These channels were combined to form a total of 4179 channels. It is noteworthy that these channels are not independent, as they share the trainable parameters $\boldsymbol{T_A}, \boldsymbol{T_B}, \boldsymbol{T_C}$. Any updates to these matrices simultaneously affect the transfer functions of all channels. To minimize crosstalk among these correlated channels as much as possible, we employed the backpropagation gradient descent algorithm of the $D^2NN$ model for joint optimization, as illustrated in **Fig. 2 a**. The transfer matrix of each channel, constructed by zero-initialized trainable coefficients, firstly transformed the input plane wave to output wavefield. Subsequently, the loss function of each channel computed the error between ground truth and field in output FOV. During each iteration, losses of all channels to be optimized were summed for parameter updates through gradient descent. To implement DMCN models with different functionalities, we adopt holographic images, focal spots, and complex vortex fields as optimization targets, and train the models using the mean squared error (MSE) loss function defined in the **Methods** part. The limited number of parameters in DMCN models makes it impractical to optimize thousands of channels simultaneously. Therefore, the channel subsets were carefully selected according to the channel selection strategy in **Supplementary Note 4** to generate 144-channel holograms, 108-channel single/double focus, 60-channel single/multi-mode OAM beam, respectively.

**Results of reconfigurable holography**

Multi-channel holography was initially taken as an example to demonstrate the effectiveness of the proposed mechanically reconfigurable strategy. By leveraging the single type of mechanical transformation, the pairwise combinations of transformations, and the integration of three transformations, the channel capacity of DMCN holography was progressively enhanced from 15 to 144. An MSE loss function with an efficiency regularization term was employed during DMCN training to optimize the image quality while controlling system efficiency. The ground truth of the quick draw, expression, and letter images used for training is presented in the **Supplementary Fig. 34**. To assess the hologram quality, the Pearson correlation coefficient (PCC) was used to evaluate the similarity between the output holograms and ground truth images, while the peak signal-to-noise ratio (PSNR) was used to analyze the noise intensity in the holograms. Detailed definitions of these metrics are provided in the **Methods** part. **Supplementary Fig. 1** presents simulation results of multi-channel quick draw holograms respectively achieved by translation (16 channels), permutation (15 channels), and rotation (16 channels). The PCCs of these holograms are 1, indicating negligible differences from the target images. **Supplementary Fig. 2** shows the simulation results of 18-channel, 24-channel and 48-channel expression holography obtained from pairwise combinations of the three mechanical configurations in **Fig. 2 b**. To balance the trade-off between image quality and system efficiency caused by the iterative spatial filtering process in the DMCN model (see **Supplementary Note 2** for details), we adopted a strategy of reducing imaging efficiency to maintain high image clarity as the channel count increases. Therefore, the target optimization efficiency $\eta_{Th}$ was respectively set to 18%, 16%, and 12% when training these models. Both the holographic images and evaluation metrics (PCC ~ 0.99, PSNR ~ 19.9dB) confirm that high image quality is maintained despite the increased number of channels.

The 144-channel letter holography model was realized by fully integrating three types of mechanical configurations, as illustrated in **Fig. 2 b**. The detailed mechanical parameters for all 144 channels are provided in **Supplementary Excel**. The model features a layer width of 400 pixels and a FOV width of 200 pixels, with each pixel measures $9\mu m * 9\mu m$. To keep high image quality, the value of $\eta_{Th}$ was further reduced to 4.5%. **Fig. 2 c** presents the phase profiles of the well-trained DMCN model. To implement the phase modulation layers of the model, we fabricated DOEs composed of 16-level photoresist relief structures through grayscale photolithography (see **Methods** for details). A partial scanning electron microscope (SEM) view of the sample is shown in **Fig. 2 c.** We further fabricated three slots to support the plug-and-play integration of DOE layers, enabling the implementation of the three mechanical transformations. **Fig. 2 d** presents the schematic diagram of the experimental optical path. The incident 532 nm laser beam was sequentially modulated by the three DOE layers and then captured by the camera at the imaging plane.

**Fig. 2 e** presents the simulated and experimental holograms of the first 6 channels together with 14 representative channels ranging from low to high PSNR. The results of remaining channels are provided in **Supplementary Fig. 4 and Supplementary Video 1**. In the simulations, the outputs of the low-PSNR channels exhibit most background noise in the central region. As the PSNR increases, the noise level decreases accordingly. Channel 46 exhibits two unexpected holes, resulting in the lowest PSNR of 19.94 dB. In contrast, the simplest character, ‘II’, achieves the highest PSNR of 27.62 dB. The PSNR of the experimental results is consistently 4~5 dB lower than that of the corresponding simulations. Despite the degradation in image quality, the letters remain clearly identifiable. **Fig. 2 f–g** shows the simulated and experimental metric histograms for all 144 channels. Simulated images attain an average PCC of 0.99 and a mean PSNR of 23.2 dB, demonstrating high imaging fidelity, while the experimental results yield a mean PCC of 0.91 and an average PSNR of 18.6 dB. Nonetheless, the distribution trends of the simulated and experimental metrics remain highly consistent. Furthermore, the channel capacity of the DMCN-based holography system can be further increased at the expense of image quality or by increasing the ratio between layer width and FOV width. The detailed trade-offs among PSNR, channel count, layer width, and FOV width ratio are analyzed in the section **"Model Scalability of DMCN"**. Our model is also capable of generating 144-channel grayscale holograms from the CIFAR-10 dataset (PCC~0.93, PSNR~19.8dB, $\eta$~4.4%), as shown in **Supplementary Fig. 33**. These reconfigurable holography results conclusively underscore the super-high information capacity of our DMCN.

**Results of reconfigurable single/double focus**

To further explore the application potential of our DMCN model, we also optimized it to achieve multi-channel single/double focus. In this scenario, both the model layer width and the FOV width were set to 600 pixels, with each pixel having a side length of 12.5 $\mu m$. During model training, the same loss function used in DMCN holography was utilized and $\eta_{Th}$ was set to 4%. The optimization target was set as focused patterns consisting of small squares centered around the target focal points. The side length of these squares could determine the full width at half maximum (FWHM) of the output Airy disks. On one hand, an excessively large side length will cause the output wavefield of the DMCN to fail in properly focusing on the target region, leading to energy dispersion. On the other hand, a too small side length will result in focal points with overly narrow FWHMs, thereby reducing focusing efficiency. Therefore, a single-layer meta-lens was chosen as a benchmark to determine the optimal side length of the target squares. **Supplementary Fig. 5** shows the phase profile of a meta-lens with the same layer width and sampling interval as the DMCN. Its focal length matches the longitudinal length of the DMCN, with a numerical aperture of 0.0062. The FWHM of its focal point is 36.25 $\mu m$, approximately 3 sampling intervals. Thus, the side length of our focused squares was set to 3 pixels. To avoid excessively low focusing efficiency, a total of 108 channels were selected for joint optimization. These channels are derived from the combination of the first six rotation channels and the other two types of channels shown in **Fig. 2 b** and their mechanical parameters are detailed in **Supplementary Excel**. Among them, 81 channels were allocated for single-focus, 27 for dual-focus. The target focusing patterns for these configurations are presented in **Supplementary Fig. 35**.

The phase profiles of the well-trained focus model are shown in **Supplementary Fig. 6**, exhibiting a more scattered distribution compared to those of the holography model. **Fig. 3 a** presents the simulated and experimental results of 6 single-focus channels and 4 dual-focus channels, selected from the total of 108 channels. Both the simulated and experimental outputs of each channel are well focused within their respective target focal regions, exhibiting high channel isolation with no observable crosstalk. **Supplementary Figs. 7-14 and Supplementary Video 2** show the focus points of other channels. To provide a more detailed illustration of the focusing results, the output fields of channels 39, 40, 51, 52, 82 and 88 are represented by their one-dimensional horizontal profiles and two-dimensional cross-sections at the focal height. **Fig. 3 b** and **Supplementary Fig. 15** present the one-dimensional profiles from the experiments and simulations, respectively. The one-dimensional curves all exhibit a maximum at the target focal point and rapidly decay to near zero as the distance from the maximum increases. The positional error of the maxima between the experimental and simulated profiles is within 8.3 μm, smaller than the side length of a DOE pixel. The simulated FWHMs are consistently 33.3 μm, while the experimental FWHM ranges from 29.2 μm to 41.6 μm. Additionally, the experimental results of channel 88 exhibit a slight unevenness in the intensities of the two focal spots. The simulated 2D field intensity distributions in **Fig. 3 c** demonstrate that the depth of focus (DOF) for these six channels is approximately 4 mm, corresponding to about 120 times the FWHM. **Fig. 3d–e** presents the histograms of evaluation metrics for both simulated and experimental results. In simulations, the mean PSNR is 42.91 dB. The FWHMs of all focal points are 33.3 $\mu m$, slightly smaller than the predesigned side length of the square patterns. For focusing efficiency, we defined the focusing energy as the total energy within three times the FWHM of the focal point and calculated an average efficiency of 3% across 108 channels. For the experiments, the mean PSNR is 42.76 dB and the FWHM centers around 35.9 $\mu m$. Experimental errors cause the metric distribution of the 108 channels to be more dispersed compared to the simulation results.

**Results of reconfigurable single/multi-mode OAM beam generation**

For multi-channel holography and focusing functionalities, the DMCN utilizes mechanical transformations to dynamically manipulate the intensity of output field. To exploit its capability of complex field manipulation, we further employed the DMCN to generate single/multi-mode OAM beams. The target single-mode OAM beams were set as Laguerre-Gaussian beams, while the target multi-mode OAM beams were formed by the superposition of multiple orders of OAM beams, which can enhance the parallelism and capacity of communication systems. The waists of the OAM beams were all set to 140 $\mu m$, and the complex light fields at the waist planes were taken as the ground truth. The DMCN requires simultaneous optimization of both the amplitude and phase of the output field to achieve complex field manipulation, which introduces an additional phase dimension compared to multi-channel holography and focusing functionalities. Consequently, the number of reconfigurable channels was reduced to 60, approximately half of the aforementioned configurations. Among these 60 channels, 51 were allocated for generating single-mode OAM beams with orders ranging from -25 to 25, while the remaining 9 were designated for producing multi-mode OAM beams formed by combining selected orders of the 51 energy-normalized beams. Specifically, the target multi-mode OAM beam of the channel 52 includes all orders of OAM beams from -25 to 25. Channels 53-56 produce beams formed by the superposition of orders -25 to -16, -15 to -6, -5 to 4, and 5 to 14, respectively. Channels 57-60 correspond to beams formed by selecting one beam every 2, 3, 4, and 5 orders from the -25 to 25 range. The optimization targets of these channels are shown in **Supplementary Fig. 36**. The DMCN model was trained using the same loss function as described above, with the output field in the loss function now represented as a complex field. The layer width of DMCN was set to 600 pixels and the side length of each pixel was 12.5 $\mu m$. To improve the purity of the OAM components in the output fields, the FOV width was reduced to 130 pixels and the efficiency hyperparameter $\eta_{Th}$ was lowered to 3.5%.

The phase profiles of the OAM model are shown in **Supplementary Fig. 6**, characterized by spiral fringe textures. **Fig. 4 a** presents the simulated intensity and phase profiles of six single-mode and four multi-mode OAM beams selected from the 60 generated channels. The order of the single-mode OAM beams in channels 1, 6, 11, 36, 41, 46 are respectively -25, -20, -15, 10, 15, 20. As the order of the OAM beam rises, both the radius of intensity rings and the tangential periodicity of phase profiles expand. The direction of the tangential phase gradient reverses when the sign of the order is inverted. Multi-mode OAM beams across four channels display distinct intensity and phase profiles, varying with their constituent orders. **Fig. 4 b** shows the intensity profiles and interference patterns of the experimentally generated single-mode OAM beams. The size of the generated donut-shaped OAM beams matches that of the simulated beams, but the intensity distributions exhibit slight non-uniformities. The interference patterns were produced by removing the beam block in **Fig. 2 d** to construct the interference path with the main modulation optical path. The OAM orders of these channels range from -1 to -7 (1th row) and 1 to 7 (2th row). The resulting patterns correctly reflect the topological charges and vortex handedness of each OAM beam. **Fig. 4 c** shows the intensity profiles of the experimentally generated multi-mode OAM beams at both the imaging plane and after 8 cm of transmission. The intensity patterns of channels 54, 56, 59, and 60 exhibit rotations, while the number of lobes in channels 57 and 58 doubles after propagation. The remaining simulated and experimental results are provided in **Supplementary Figs. 16-17 and Supplementary Video 3**.

The correlation coefficient (CC) matrix of the 51 simulated OAM beams is shown in **Fig. 4 d**. The diagonal elements of this matrix represent the correlation between the output and target beams, with values close to 1, indicating the high purity of each OAM beam. The other elements, with values near 0, indicate that the generated OAM beams are uncorrelated, highlighting strong channel orthogonality and negligible crosstalk. **Fig. 4 d** also illustrates the histogram of the 60-channel transmission efficiency, where most channel efficiencies are concentrated around 2.9%. **Fig. 4 e** displays the simulated OAM spectra of the selected 4-channel multi-mode OAM beams. The definition formula of the spectral coefficients is detailed in the **Methods**. Their OAM spectra indicate that each order component has a highly uniform intensity distribution, with coefficients ranging from 0.95 to 1. Furthermore, the order of the generated OAM beams can be enhanced. Two examples of high-order OAM beam generations are provided in **Supplementary Figs. 31-32.**

**The model scalability of DMCN**

To analyze the model scalability of DMCN, we propose an equivalent connection law under the condition of large modulation pixel size (side length $\geq 1.22\lambda$). This law is detailed demonstrated in the **Methods** part and derived in the **Supplementary Note 1**. It assesses the performance similarity among different DMCN models by analyzing the connectivity between adjacent modulation layers. DMCN models with the same inter-layer connectivity $\alpha$ are defined as equivalent-connection models, which can be further classified into identical-connection models (with same N) and similar-connection models (with different N). The identical-connection models exhibit exactly the same functionality when loaded with the same set of phase modulation coefficients. **Fig. 5 a** presents the output optical fields of identical connection models corresponding to the 144-channel holography model and the 60-channel OAM model. These models are loaded with the phase modulation coefficients derived from the aforementioned holography and OAM models. The equivalent models, with the pixel size $p$ scaled by a factor of $\sqrt{2}$ and the layer spacing $d$ scaled by a factor of 2, yield output optical fields with quality comparable to that of the original models. However, when $p$ is reduced by a factor of 10 and $d$ by a factor of 100, the output field quality drops significantly. This is because the $\beta$ values for both models fall below the valid approximation range. When the wavelength of a DMCN model is doubled or halved, same performance can still be achieved by proportionally reducing or increasing the layer spacing. Additionally, **Supplementary Fig. 24** demonstrates the bandwidth of DMCN models implemented with DOE based on the identical connection principle (i.e., in addition to scaling wavelength and layer spacing, the changes in phase profiles of DOEs caused by wavelength variation were also taken into account). After compensating for wavelength variation by adjusting layer spacing, the models are capable of producing satisfactory output fields across a wide wavelength range of $450\sim600$ nm. For one thing, the identical connection principle decouples the DMCN model weights from hardware platforms, allowing seamless migration across devices with different parameters. For another, it expands the bandwidth of the model optimized at a single wavelength. Furthermore, since the inter-layer connectivity of a DMCN model determines its performance[44,45], similar connection models tend to exhibit comparable performance on the same task. This means that optimal hyperparameter settings for a small-scale model (with small N) can be transferred to a large-scale model to accelerate the training process. Given that our DMCN model is described and optimized based on the $D^2NN$ framework, these properties are also applicable to $D^2NN$ models in other application scenarios.

We selected 576 channels as shown in **Supplementary Excel** and used reconfigurable holography as an example to discuss the relationship between layer width, FOV width, channel number, and PSNR. The average system efficiencies of all simulated models in this section are maintained around 4.5%. To keep the connectivity between adjacent layers, the unit distance interval scales proportionally as the layer width changes. The relationship between the channel count and PSNR is shown in the **Fig. 5 b**. The layer width and FOV width were fixed at 400 pixels and 200 pixels. The image PSNR decreases as the number of optimized channels increases. At 576 channels, the average PSNR and PCC are respectively 16.5 dB and 0.892. Despite the significant increase in background noise, the letters remain distinguishable. Using a PCC threshold of 0.9, our DMCN model can generate up to 576-channel letter holography. **Fig. 5 c** shows an approximately linear relationship with a slope of about 0.58 between layer width and FOV width, when the number of channels and PSNR are fixed. This indicates that increasing the layer width allows for an enlargement of the output FOV without compromising the model capacity and imaging quality. In the **Fig. 5 d** and **Fig. 5 e**, the FOV width is set as 200 pixels. We only adjusted the layer width to alter the ratio of layer width to FOV width and analyzed its impact on channel capacity and quality. Enlarging the layer width leads to the improvement in image PSNR, with the number of channels kept constant. If the layer width is below 300 pixels, the improvement in image quality is pronounced. However, once it exceeds 300 pixels, the rate of improvement slows as the image quality has already reached a high level. An expansion in layer width also leads to nearly linear growth in the number of channels while preserving a mean PSNR of 23 dB. Each 100-pixel increase of layer width yields approximately 44 additional channels. Therefore, the information capacity and channel quality of DMCN can be effectively improved by increasing the ratio of layer width to FOV width.

**Discussion**

In this article, a mechanically-enabled DMCN model is proposed to produce high-quality 144-channel holograms, 108-channel single/double focus, and 60-channel single/multi-mode OAM beam. The successful implementation of these functions demonstrates that DMCN can manipulate both real-valued field intensities and complex-valued fields, offering exceptionally high information capacity and flexibility. The channel capacity and quality can be further enhanced by increasing the layer width and the FOV width ratio. At the theoretical level, an equivalent connection law is proposed to guide DMCN models in achieving hardware decoupling, bandwidth expansion, and hyperparameter optimization. Our strategy is highly versatile, as it allows the implementation of various reconfigurable functionalities by simply modifying the target wavefields. This offers a novel approach to enhancing the capacity of optical information systems. Since the DMCN model only requires simple phase-only modulation units, it can also be implemented using other common devices, such as metasurfaces and LCs (for detailed analysis, see **Supplementary Note 5**). It can also be extended to other spectral bands, including the terahertz and microwave regions. The reconfigurable channels of DMCN are solely formed by mechanical structural variations, making it easy to integrate with other multiplexing strategies such as polarization or wavelength multiplexing to further enhance capacity. Furthermore, the thousands of potential channels available for optimization provide flexible combinations for implementing DMCN. **Supplementary Note 6** discusses the results of diverse combinations in detail.

The three-layer DMCN models demonstrated in this paper possess 4179 potential channels, and expanding the model to five layers will boost the number of potential channels to approximately 36 million. However, only a few hundred of these channels will be ultimately optimized, meaning that effective information is allocated to fewer than 3% of the potential channels. As a result, retrieving information from the output plane becomes impossible if the correctly configured channels are unknown. Thus, the DMCN has strong capabilities of encryption and offers promise for realizing novel holographic encryption systems[35,46]. The reconfigurable focusing system realized by DMCN can accurately switch focal positions and numbers through simple mechanical transformations. Such highly flexible and customizable focusing functionalities will provide a new platform for applications like lithography, microscopy, and optical sensing[41, 47]. By defining appropriate loss functions, the DMCN can dynamically generate OAM beams of any order or multi-mode OAM beams with arbitrary components and relative intensities. Furthermore, the DMCN can manipulate other complex fields, such as OAM combs, perfect vortex beams, Bessel beams, and Airy beams. This versatility opens up possibilities for generating specialized beams, performing optical beam operations, and enhancing the capacity of optical communication systems[48-50]. In summary, our DMCN model holds promising prospects for applications in optical computing, optical communication, optical image processing, and optical sensing.

## Methods

### Forward propagation model of DMCN

The DMCN comprises three phase modulation layers and an output layer. The modulation layers transform the incident wavefield according to the following formula:

$$u_l^t(x,y) = u_l^i(x,y)t_l(x,y)$$
$$t_l(x,y) = \exp(j2\pi Sigmoid(\phi_l(x,y))) \tag{1}$$

where $u_l^i(x,y)$ and $u_l^t(x,y)$ respectively represent incident and transmitted wave of the $l$-th layer ($1 \leq l < 3$). $t_l(x,y)$ represents the transmission coefficient of $l$-th layer and serves as learnable parameters.

Propagation procedure between two adjacent layers can be expressed using the Rayleigh Sommerfeld convolution (RSC)[51] formula as follows:

$$u_{l+1}^i(x,y) = u_l^t(x,y) * w(x,y,d_l)$$
$$w(x,y,d_l) = \frac{d_l}{r^2}\left(\frac{1}{2\pi r} + \frac{1}{j\lambda}\right)exp\left(j\frac{2\pi r}{\lambda}\right)$$
$$r = \sqrt{x^2 + y^2 + d_l^2} \tag{2}$$

Here, $d_l$ $(1 \leq l < 3)$ represents the distance between two layers, and $w(x,y,d_l)$ is the convolution operator that captures the diffraction connections.

Although the RSC method can accurately depict the light diffraction process in DMCN, the computational overhead caused by large-kernel convolutions makes it impractical for model training. Therefore, we only utilized this equation for numerical validation. Instead, a lower-computing and faster algorithm known as angular spectrum method (ASM)[51] was employed for model training and inference, which can be expressed as follows:

$$u_{l+1}^i(x,y) = IFFT(FFT(u_l^t(x,y))H(f_x,f_y,d_l))$$
$$H(f_x,f_y,d_l) = \exp\left(\frac{j2\pi d_l}{\lambda}\sqrt{1-(\lambda f_x)^2-(\lambda f_y)^2}\right) \tag{3}$$

where $H(f_x,f_y,d_l)$ is the spectrum transfer function, $f_x$ and $f_y$ are spatial frequencies along x and y axis, respectively.

**The mechanically reconfigurable non-orthogonal channels of DMCN**

The multiplexing channels of DMCN are realized through the combination of three types of mechanical operations: rotation, permutation, and translation. To theoretically demonstrate the reconfigurable principle, we derived the output wavefield of DMCN in compact form under the plane wave illumination by discretizing equations (1) and (2).

$$Vec(\boldsymbol{O_c}) = \boldsymbol{W}(d_3)Diag\big(Rot(\boldsymbol{T_3}, \Omega_2)\big)\boldsymbol{W}(d_2)Diag\big(Rot(\boldsymbol{T_2}, \Omega_1)\big)\boldsymbol{W}(d_1)Diag(\boldsymbol{T_1})Vec(\boldsymbol{J}) \tag{4}$$

Here, $\boldsymbol{J}$ is the all-one matrix that represents the input plane wave, $\boldsymbol{T_l}$ is the transmission coefficient matrix and $\boldsymbol{W}(d_l)$ is the diffraction matrix composed of convolutional kernels. $Vec(\cdot)$ operator flattens the matrix into column vector. $Diag(\cdot)$ operator transforms all elements of the matrix into a diagonal vector, thereby forming a diagonal matrix. $Rot(\cdot,\ \cdot)$ operator rotates the matrix counterclockwise by a certain angle.

From equation (4), the transfer matrix of DMCN can be further derived as follows:

$$\boldsymbol{A_c} = \boldsymbol{W}(d_3)Diag\big(Rot(\boldsymbol{T_3}, \Omega_2)\big)\boldsymbol{W}(d_2)Diag\big(Rot(\boldsymbol{T_2}, \Omega_1)\big)\boldsymbol{W}(d_1)Diag(\boldsymbol{T_1}) \tag{5}$$

This transfer matrix is solely determined by the system parameters and is independent of the input. The parameters $\Omega_1$ and $\Omega_2$ are adjusted through rotation, while $d_1$, $d_2$ and $d_3$ are controlled by translation. The permutation operation alters the order of the three transmission layers, thereby affecting the values of $\boldsymbol{T_1}$, $\boldsymbol{T_2}$ and $\boldsymbol{T_3}$. Each operation can modify the transfer matrix, generating a new non-orthogonal channel. These channels are capable of performing different tasks with the same input.

**Loss function and evaluation criteria**

We employed the following loss function to train the DMCN.

$$\mathcal{L}_{DMCN} = \mathcal{L}_{MSE} + \gamma_{Eff}\mathcal{L}_{Eff} \tag{6}$$

Here, $\mathcal{L}_{MSE}$ quantifies the error between the output and target wavefields and is defined as follows:

$$\mathcal{L}_{MSE} = \frac{1}{N}\sum_{c=1}^{N} E[|\sigma_1 G_c(\mathrm{x},\mathrm{y}) - \sigma_2 K_c(\mathrm{x},\mathrm{y})|^2]$$

$$\sigma_1 = \frac{1}{\sum_{(x,y)\in FOV_O}|G_c(\mathrm{x},\mathrm{y})|^2}$$

$$\sigma_2 = \sigma_1 \frac{\sum_{(x,y)\in FOV_O} G_c(\mathrm{x},\mathrm{y})K_c(\mathrm{x},\mathrm{y})^*}{\sum_{(x,y)\in FOV_O}|K_c(\mathrm{x},\mathrm{y})|^2} \tag{7}$$

In these equations, $E[\cdot]$ is the average operator, while $G_c(x,y)$ and $K_c(x,y)$ stand the training target and the model output of the c-th channel, respectively. $N$ is the number of channels. $\sigma_1$ and $\sigma_2$ are normalization factors. For Hologram and focus applications, $K_c(\mathrm{x},\mathrm{y})$ is defined as the intensity of output wavefield $|O_c(x,y)|^2$. For OAM application, $K_c(\mathrm{x},\mathrm{y})$ is defined as the complex output wavefield $O_c(x,y)$.

The term $\mathcal{L}_{Eff}$ is a regularization factor designed to ensure system efficiency, defined as:

$$\mathcal{L}_{Eff} = \frac{1}{N}\sum_{c=1}^{N} \mathcal{L}_{subEff}^{c}$$

$$\mathcal{L}_{subEff}^{c} = \begin{cases} -\ln\left(\frac{\eta_c}{\eta_{Th}}\right), & \eta_c < \eta_{Th} \\ 0, & \eta_c \geq \eta_{Th} \end{cases}$$

$$\eta_c = \frac{\sum_{(x,y)\in ROI}|O_c(x,y)|^2}{\sum_{(x,y)}|J(x,y)|^2} \tag{8}$$

where $\eta_c$ represents the efficiency of the c-th channel, $\eta_{Th}$ is a hyperparameter that specifies the target efficiency, and $J(x,y)$ denotes the input plane wave. The region of interest (ROI) corresponds to $FOV_O$ for hologram and OAM applications, and to the target focus region for focus applications. $\gamma_{Eff}$ is the weight coefficient for the efficiency regulation term, which was set as 0.01 for all the models.

For hologram and focus evaluation, we further took the PCC and PSNR as metrics, which are defined as follows:

$$PCC(G(x,y), O(x,y)) = \frac{\sum_{(x,y)}(\mathrm{G(x,y)} - \mathrm{E[G(x,y)]})(\mathrm{O(x,y)} - \mathrm{E[O(x,y)]})}{\sqrt{\sum_{(x,y)}|\mathrm{G(x,y)} - \mathrm{E[G(x,y)]}|^2 \sum_{(x,y)}|\mathrm{O(x,y)} - \mathrm{E[O(x,y)]}|^2}} \tag{9}$$

$$PSNR(G(x,y), O(x,y)) = 10log_{10}\frac{1}{E\left[\left|\tilde{G}(x,y) - \tilde{O}(x,y)\right|^2\right]} \tag{10}$$

where $\tilde{\ }$ is the max-normalization operator.

For OAM evaluation, CC was determined as follows to assess model performance:

$$CC(G(x,y),O(x,y)) = \frac{|\sum_{(x,y)} G(x,y)O(x,y)^*|}{\sqrt{\sum_{(x,y)}|G(x,y)|^2 \sum_{(x,y)}|O(x,y)|^2}} \tag{11}$$

Lastly, the normalized OAM coefficients defined below were employed to analyze the OAM components contained in generated combs:

$$C_l = \int_0^\infty \left|\int_0^{2\pi} O(r,\varphi)\exp(-il\varphi)\,d\varphi\right|^2 rdr$$

$$\widetilde{C_l} = \frac{C_l}{\max\limits_{-\infty<q<\infty} C_q} \tag{12}$$

where $O(r,\varphi)$ is the output field in spherical coordinate.

**The equivalent connection law of DMCN**

**Figure 5 a** shows a side-view schematic of two adjacent square modulation layers. we define the connectivity $\alpha$ as the ratio of the side length in the next layer covered by the maximum diffraction angle of the edge modulation unit in the preceding layer to the total side length.

$$\alpha \approx \frac{\lambda d}{2Np^2} \propto \frac{\lambda d}{Np^2}, \beta \geq \frac{\sqrt{5}}{2} \approx 1.12 \tag{13}$$

Here, $p$ denotes the side length of each modulation pixel, and the ratio of $p$ to the operating wavelength $\lambda$ is defined as $\beta$, which is typically greater than 10 for DOE in the visible spectrum. $d$ represents the spacing between adjacent layers, and $N$ is the number of modulation units along one side of the square. $\alpha < 1$ corresponds to under-connection, $\alpha = 1$ to critical connection, and $\alpha > 1$ to over-connection.

Specifically, when the parameter $N$ is fixed and $\beta$ further increases, the discretized angular spectrum diffraction transfer kernel $H(\Delta f_x, \Delta f_y, d)$ can be uniquely determined by the connectivity $\alpha$. This one-to-one mapping $f$ can be written as follows:

$$f: \alpha \propto \frac{\lambda d}{p^2} \rightarrow H(\Delta f_x, \Delta f_y, d), \beta \geq \sqrt{5}\delta \approx 2.24\delta \tag{14}$$

Where $\delta$ denotes the upsampling factor applied to each modulation pixel. This indicates that, under the above conditions, two sets of adjacent modulation layers with identical connectivity exhibit identical diffraction transfer behavior. Two DMCN models with different parameters are defined as equivalent connection models if, in any mechanical configuration, each pair of adjacent modulation layers has the same connectivity (the output layer is viewed as a special layer characterized by a unit modulation coefficient). Specifically, if two equivalent connection DMCN models share the same $N$, they are referred to as identical connection DMCN models. Otherwise, if $N$ differs, they are called similar connection DMCN models.

**Numerical and Experimental implementations**

Our numerical DMCN models were implemented using Python and PyTorch on a workstation equipped with a Xeon® Gold 6430 CPU and a GeForce RTX 4090 GPU. To reduce numerical computation errors in the holography model, we increased the spectral sampling rate of the FFT by zero-padding to four times the original signal length ($M = 4N,\ \delta = 1$). Building upon this, we further minimized errors in the focus and OAM models by enhancing the spatial sampling rate through two-fold pixel-wise upsampling ($M = 4N,\ \delta = 2$). During each training iteration, the full batch containing all channel data pairs was fed into the model. And the Adam optimizer with a learning rate of 0.1 was used for backpropagation (BP) to update the transmission coefficients of the DMCN. The training procedure last around 10~15 minutes, involving 1500 epochs to ensure sufficient optimization.

For the experimental setup, A diode-pumped solid-state laser (LDM56, λ = 532 nm) was employed as the light source. The incident beam was directed onto a transmission system comprising three diffractive optical element (DOE) layers. And the output fields were recorded by a CMOS camera (MUS500C.G) with a resolution of 2464 × 2056 pixels and a pixel size of 3.45 μm. The fabrication process for the DOE started with ultrasonic cleaning of a fused silica substrate sequentially in acetone, isopropyl alcohol (IPA), and deionized water (DIW). The substrate was then pre-baked on a hot plate at 120 °C for 2 min to promote photoresist adhesion. After cooling, a positive photoresist (AZ 5214) was spin-coated to a thickness of 1.68 μm and soft-baked at 105 °C for 60 s. The refractive index of the resist was measured by ellipsometry (Horiba Uvisel Plus) as 1.46. The desired DOE structures were defined using grayscale UV lithography (TTT-07-UV Litho-ACA), followed by development in AZ MIF300 for 30 s and rinsing in DIW for 30 s. Each DOE was designed with 16-level phase quantization, and the quantization error analysis is provided in **Supplementary Fig. 22 a**.

During the experiments, the DOE components were mounted in CNC-machined slot structures, each slot containing three grooves to enable mechanical switching of DOE layers. The slots were integrated with five-axis stages, combining two-axis rotation/pitch control with a three-axis differential micrometer translation stage (resolution 1 μm). This configuration allowed precise alignment of the second and third DOE layers and rapid replacement of components once interlayer distances were set. The actual implementations of the optical path and the slot structure are shown in **Supplementary Fig. 26**. To minimize alignment errors, double-nested lens phase patterns were employed for calibration, enabling far-field diffraction–based alignment[52]. The alignment tolerance for holographic functions was the strictest, with lateral tolerance of ±4.5 μm and longitudinal tolerance of ±0.5 mm (see **Supplementary Note 3** for details).

## Disclosures
The authors declare no competing interests.

## Code and Data Availability
The deep learning models reported in this work used standard libraries and scripts that are publicly available in PyTorch. All the data and methods needed to evaluate the conclusions of this work are presented in the main text and Supplementary Materials. Additional data can be requested from the corresponding author.

## Acknowledgements
The authors acknowledge the financial support by the National Natural Science Foundation of China (Grants No.61991423).

**References**

1. Arbabi A, Horie Y, Bagheri M, Faraon A. Dielectric metasurfaces for complete control of phase and polarization with subwavelength spatial resolution and high transmission. *Nat Nanotechnol* **10**, 937-943 (2015).

2. Wang B*, et al.* Visible-Frequency Dielectric Metasurfaces for Multiwavelength Achromatic and Highly Dispersive Holograms. *Nano Lett* **16**, 5235-5240 (2016).

3. Hu Y*, et al.* Electrically Tunable Multifunctional Polarization-Dependent Metasurfaces Integrated with Liquid Crystals in the Visible Region. *Nano Lett* **21**, 4554-4562 (2021).

4. Li P*, et al.* Multidimensional Multiplexing Liquid Crystal Holograms. *Laser & Photonics Reviews* **18**, 2400088 (2024).

5. Wang J*, et al.* The Polarization Multiplexing Image with a Single Diffractive Optical Element. *IEEE Photonics Journal* **9**, 1-8 (2017).

6. He J, Ye J, Wang X, Kan Q, Zhang Y. A broadband terahertz ultrathin multi-focus lens. *Sci Rep* **6**, 28800 (2016).

7. Khonina SN, Kazanskiy NL, Butt MA. Exploring Diffractive Optical Elements and Their Potential in Free Space Optics and imaging‐A Comprehensive Review. *Laser & Photonics Reviews* **18**, 2400377 (2024).

8. Luo Z, Li Y, Semmen J, Rao Y, Wu ST. Achromatic diffractive liquid-crystal optics for virtual reality displays. *Light Sci Appl* **12**, 230 (2023).

9. Liu Z*, et al.* Broadband, Low‐Crosstalk, and Massive‐Channels OAM Modes De/Multiplexing Based on Optical Diffraction Neural Network. *Laser & Photonics Reviews* **17**, 2200536 (2023).

10. Chong M-Z*, et al.* Generation of polarization-multiplexed terahertz orbital angular momentum combs via all-silicon metasurfaces. *Light: Advanced Manufacturing* **5**, 400-409 (2024).

11. Liu SJ, Chen P, Ge SJ, Zhu L, Zhang YH, Lu YQ. 3D Engineering of Orbital Angular Momentum Beams via Liquid‐Crystal Geometric Phase. *Laser & Photonics Reviews* **16**, 2200118 (2022).

12. Hu ZY*, et al.* Broad-Bandwidth Micro-Diffractive Optical Elements. *Laser & Photonics Reviews* **16**, 2100537 (2021).

13. Lin X*, et al.* All-optical machine learning using diffractive deep neural networks. *Science* **361**, 1004-1008 (2018).

14. Ding X*, et al.* Metasurface-Based Optical Logic Operators Driven by Diffractive Neural Networks. *Adv Mater* **36**, 2470071 (2024).

15. Zhao Z*, et al.* Deep learning-enabled compact optical trigonometric operator with metasurface. *PhotoniX* **3**, 15 (2022).

16. Liang X*, et al.* All-Optical Multi-Order Multiplexing Differentiation Based on Dynamic Liquid Crystals. *Laser & Photonics Reviews* **18**, 2400032 (2024).

17. Huang YW*, et al.* Aluminum plasmonic multicolor meta-hologram. *Nano Lett* **15**, 3122-3127 (2015).

18. Luo X*, et al.* Metasurface-enabled on-chip multiplexed diffractive neural networks in the visible. *Light Sci Appl* **11**, 158 (2022).

19. Xu H-X*, et al.* Completely Spin-Decoupled Dual-Phase Hybrid Metasurfaces for Arbitrary Wavefront Control. *ACS Photonics* **6**, 211-220 (2018).

20. Balthasar Mueller JP, Rubin NA, Devlin RC, Groever B, Capasso F. Metasurface Polarization Optics: Independent Phase Control of Arbitrary Orthogonal States of Polarization. *Phys Rev Lett* **118**, 113901 (2017).

21. Zeitner UD, Schnabel B, Kley E-B, Wyrowski F. Polarization multiplexing of diffractive elements with metal-stripe grating pixels. *Applied optics* **38**, 2177-2181 (1999).

22. Ren H*, et al.* Metasurface orbital angular momentum holography. *Nat Commun* **10**, 2986 (2019).

23. Chen K*, et al.* Directional Janus Metasurface. *Adv Mater* **32**, e1906352 (2020).

24. Xiong B*, et al.* Breaking the limitation of polarization multiplexing in optical metasurfaces with engineered noise. *Science* **379**, 294-299 (2023).

25. Wang Y, Pang C, Qi J. 3D Reconfigurable Vectorial Holography via a Dual-Layer Hybrid Metasurface Device. *Laser & Photonics Reviews* **18**, 2300832 (2023).

26. Wang Y, Yu A, Cheng Y, Qi J. Matrix Diffractive Deep Neural Networks Merging Polarization into Meta-Devices. *Laser & Photonics Reviews* **18**, 2300903 (2023).

27. Wang J*, et al.* Unlocking ultra-high holographic information capacity through nonorthogonal polarization multiplexing. *Nat Commun* **15**, 6284 (2024).

28. Liu C*, et al.* A programmable diffractive deep neural network based on a digital-coding metasurface array. *Nature Electronics* **5**, 113-122 (2022).

29. Li L*, et al.* Electromagnetic reprogrammable coding-metasurface holograms. *Nat Commun* **8**, 197 (2017).

30. Sandford O'Neill J*, et al.* 3D Switchable Diffractive Optical Elements Fabricated with Two-Photon Polymerization. *Advanced Optical Materials* **10**, 2102446 (2022).

31. Sun S, Li J, Li X, Zhao X, Li K, Chen L. Dynamic 3D metasurface holography via cascaded polymer dispersed liquid crystal. *Microsyst Nanoeng* **10**, 203 (2024).

32. Zhang XG*, et al.* An optically driven digital metasurface for programming electromagnetic functions. *Nature Electronics* **3**, 165-171 (2020).

33. Chen B*, et al.* Directional terahertz holography with thermally active Janus metasurface. *Light Sci Appl* **12**, 136 (2023).

34. He G, Qian C, Jia Y, Fan Z, Wang H, Chen H. Twisted Metasurfaces for On-Demand Focusing Localization. *Advanced Optical Materials*, 2401933 (2024).

35. Georgi P*, et al.* Optical secret sharing with cascaded metasurface holography. *Science Advances* **7**, eabf9718 (2021).

36. Wei Q*, et al.* Rotational Multiplexing Method Based on Cascaded Metasurface Holography. *Advanced Optical Materials* **10**, 2102166 (2022).

37. Wang Z*, et al.* Cascaded liquid crystal holography for optical encryption. *Chinese Optics Letters* **21**, 120003 (2023).

38. Bernet S. Tunable two-element diffractive lenses with optimal bandwidth utilization. *Opt Express* **30**, 38717-38726 (2022).

39. Gharbi Ghebjagh S, Behrens A, Fesser P, Sinzinger S. Rotationally tunable multi-focal diffractive moire lenses. *Appl Opt* **60**, 5145-5152 (2021).

40. Fan Z*, et al.* Holographic multiplexing metasurface with twisted diffractive neural network. *Nat Commun* **15**, 9416 (2024).

41. Wang X*, et al.* 3D Nanolithography via Holographic Multi-Focus Metalens. *Laser & Photonics Reviews* **18**, 2400181 (2024).

42. Wu Y*, et al.* Tbps wide-field parallel optical wireless communications based on a metasurface beam splitter. *Nat Commun* **15**, 7744 (2024).

43. Li W, Yu Q, Qiu JH, Qi J. Intelligent wireless power transfer via a 2-bit compact reconfigurable transmissive-metasurface-based router. *Nat Commun* **15**, 2807 (2024).

44. Zheng M, Shi L, Zi J. Optimize performance of a diffractive neural network by controlling the Fresnel number. *Photonics Research* **10**, 2667-2676 (2022).

45. Chen H*, et al.* Diffractive Deep Neural Networks at Visible Wavelengths. *Engineering* **7**, 1483-1491 (2021).

46. Guo X*, et al.* Stokes meta-hologram toward optical cryptography. *Nat Commun* **13**, 6687 (2022).

47. Suresh SA*, et al.* All-Dielectric Meta-Microlens-Array Confocal Fluorescence Microscopy. *Laser & Photonics Reviews*, 2401314 (2024).

48. Deng M*, et al.* Broadband angular spectrum differentiation using dielectric metasurfaces. *Nat Commun* **15**, 2237 (2024).

49. Wen J*, et al.* All-Dielectric Synthetic-Phase Metasurfaces Generating Practical Airy Beams. *ACS Nano* **15**, 1030-1038 (2021).

50. Wang J*, et al.* Terabit free-space data transmission employing orbital angular momentum multiplexing. *Nature Photonics* **6**, 488-496 (2012).

51. Zhang W, Zhang H, Sheppard CJR, Jin G. Analysis of numerical diffraction calculation methods: from the perspective of phase space optics and the sampling theorem. *J Opt Soc Am A Opt Image Sci Vis* **37**, 1748-1766 (2020).

52. Ghahremani M, McClung A, Mirzapourbeinekalaye B, Arbabi A. 3D alignment of distant patterns with deep-subwavelength precision using metasurfaces. *Nat Commun* **15**, 8864 (2024).

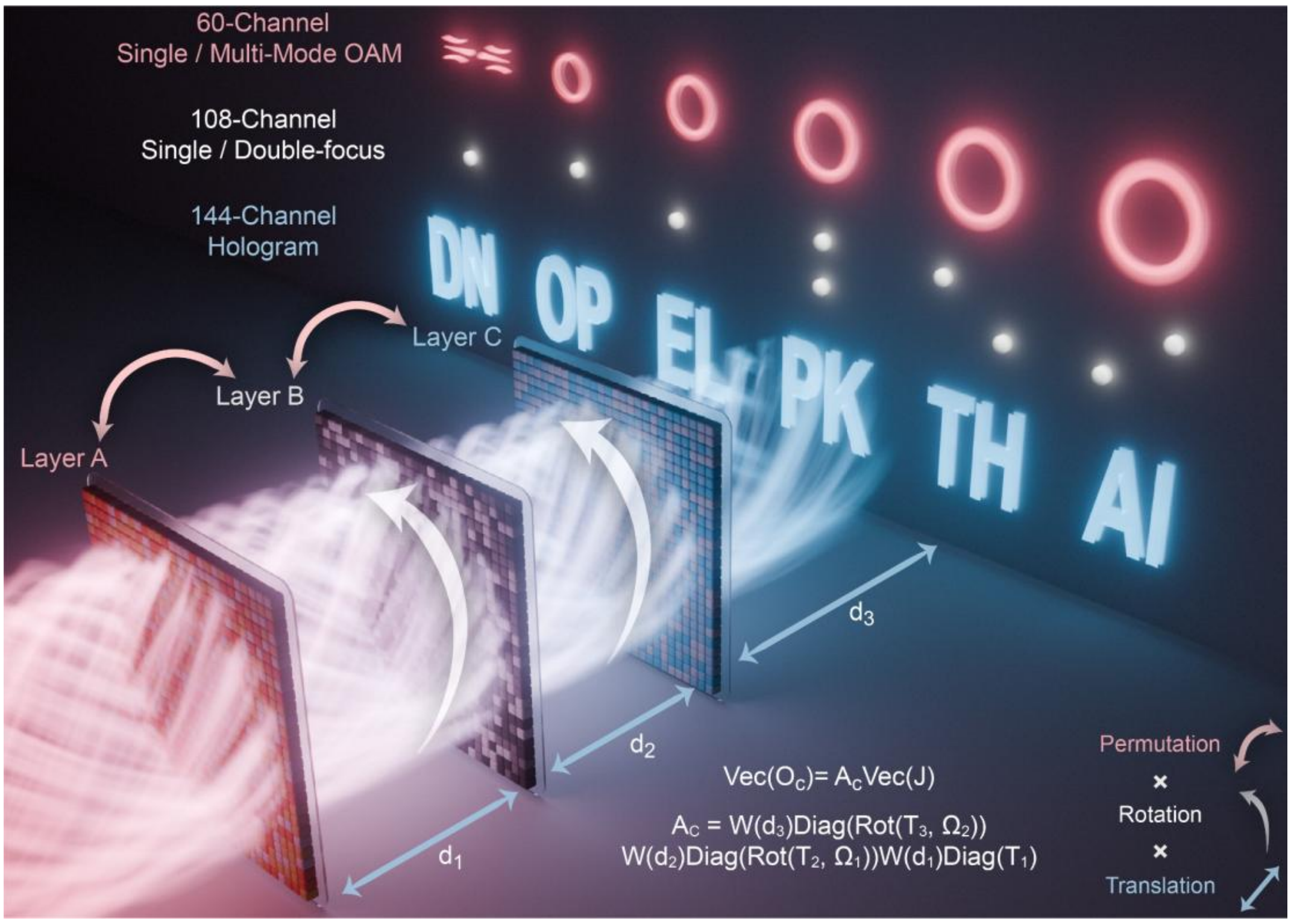

**Fig. 1 | Schematic of DMCN.** The DMCN consists of three phase-only modulation layers. Using the D[2]NN model to describe the wave propagation process, we derived the channel system matrix $\boldsymbol{A_c}$ (see **Methods** for details). Each mechanical operation introduces a variation to the formula of $\boldsymbol{A_c}$, thereby generating new channels. Specifically, permutation alters the correspondence between $\boldsymbol{T_1}, \boldsymbol{T_2}, \boldsymbol{T_3}$ and layers A, B, and C. Rotation modifies the values of $\Omega_1$ and $\Omega_2$. Translation changes the values of $d_1$, $d_2$, $d_3$.

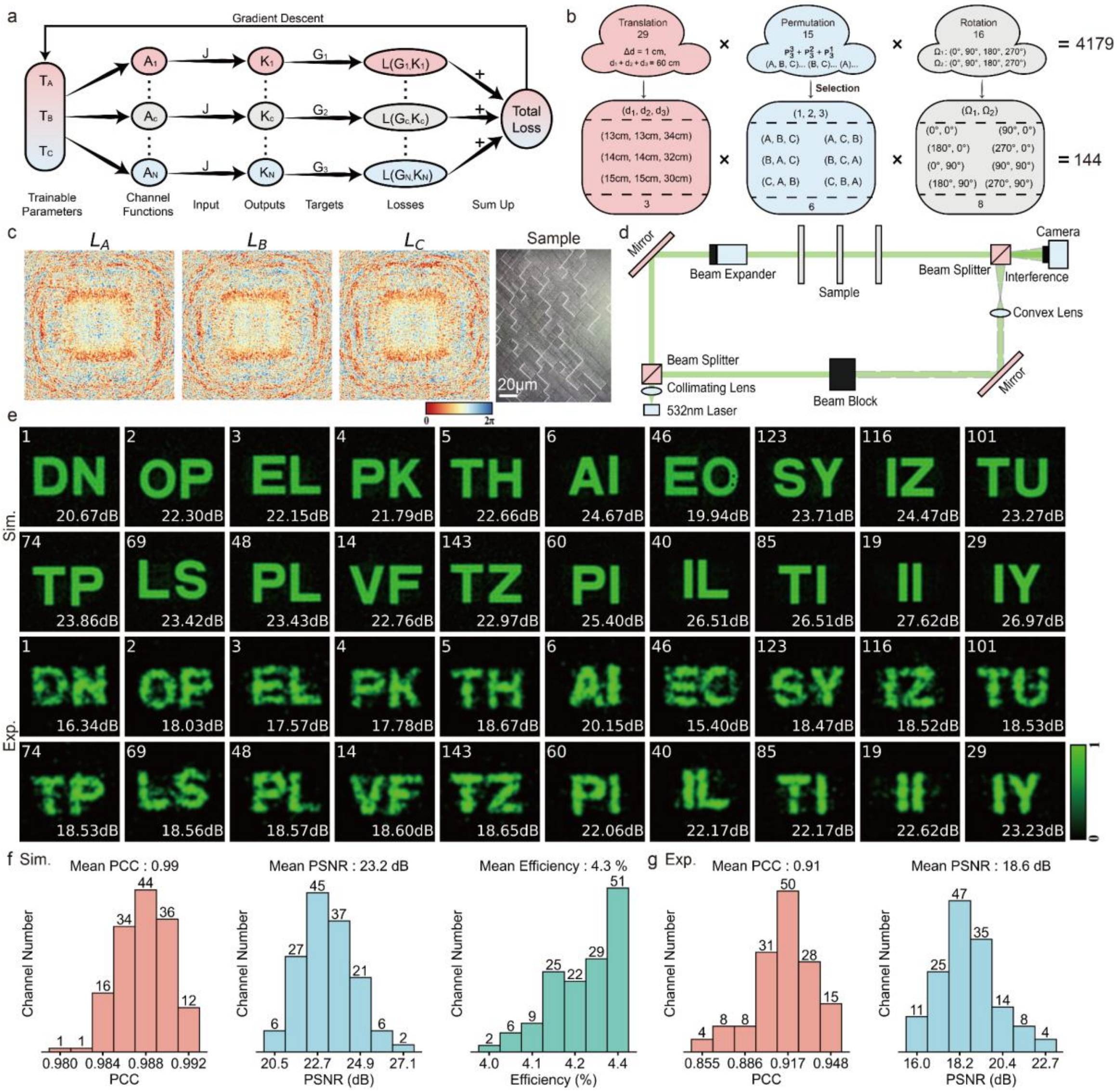

**Fig. 2 | Multi-channel optimization algorithms and results of mechanically reconfigurable holography. a** The joint optimization procedure of multiple channels. $\boldsymbol{T}_{\mathbf{A}}$, $\boldsymbol{T}_{\mathbf{B}}$, $\boldsymbol{T}_{\mathbf{C}}$ are set as trainable parameters. For holography and focus, $K_c(\mathrm{x,y}) = |O_c(\mathrm{x,y})|^2$. For OAM generation, $K_c(\mathrm{x,y}) = O_c(\mathrm{x,y})$. **b** Potential and selected channels. The permutation, rotation, and translation respectively generate 16, 15, and 29 channels, being combined to generate 4179 channels ($P_m^n = \frac{m!}{(m-n)!}$). 144 channels to be optimized are carefully selected from them. **c** Three-layer phase profiles of holographic DMCN and a partial scanning electron microscope (SEM) view of the fabricated DOE. **d** Schematic diagram of the experimental optical setup. The beam block will be removed for interferometric measurement of the topological charge. **e** Simulated and experimental results of 20 representative channels from the 144-channel holograms. **f-g** Metric histograms of the simulated and experimental 144-channel holograms.

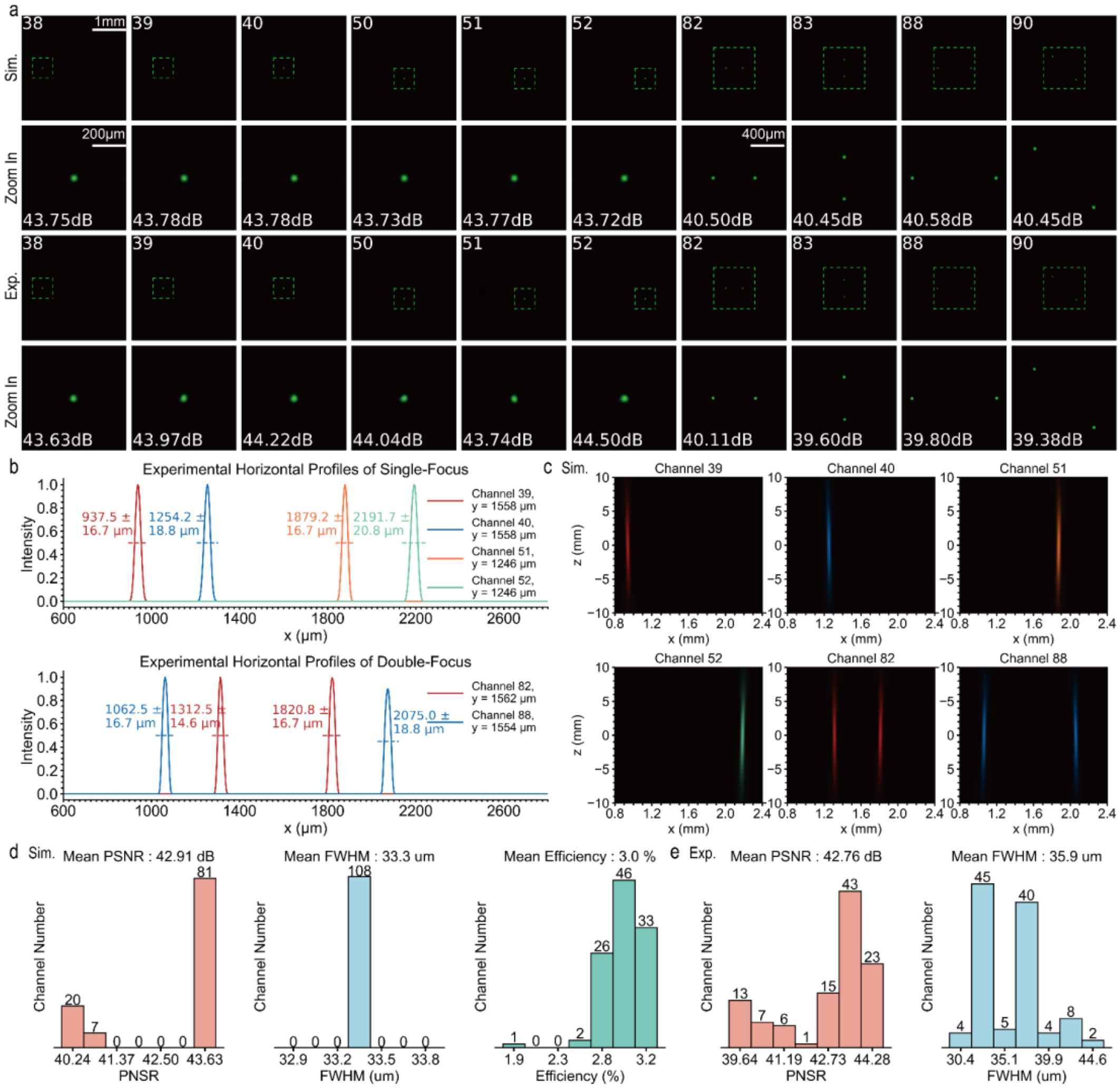

**Fig. 3 | Results of mechanically reconfigurable single/double focus. a** Simulated and experimental results of 10 representative channels from the 108-channel focuses. **b** Experimental one-dimensional horizontal profiles of the focal heights at the focal plane for channels 39, 40, 51, 52, 82, 88. **c** Simulated two-dimensioanl horizontal cross-sections of the focal heights for the aforementioned 6 channels. Focal plane locates at Z = 0 $mm$. **d-e** Metric histograms of the simulated and experimental 108-channel focuses.

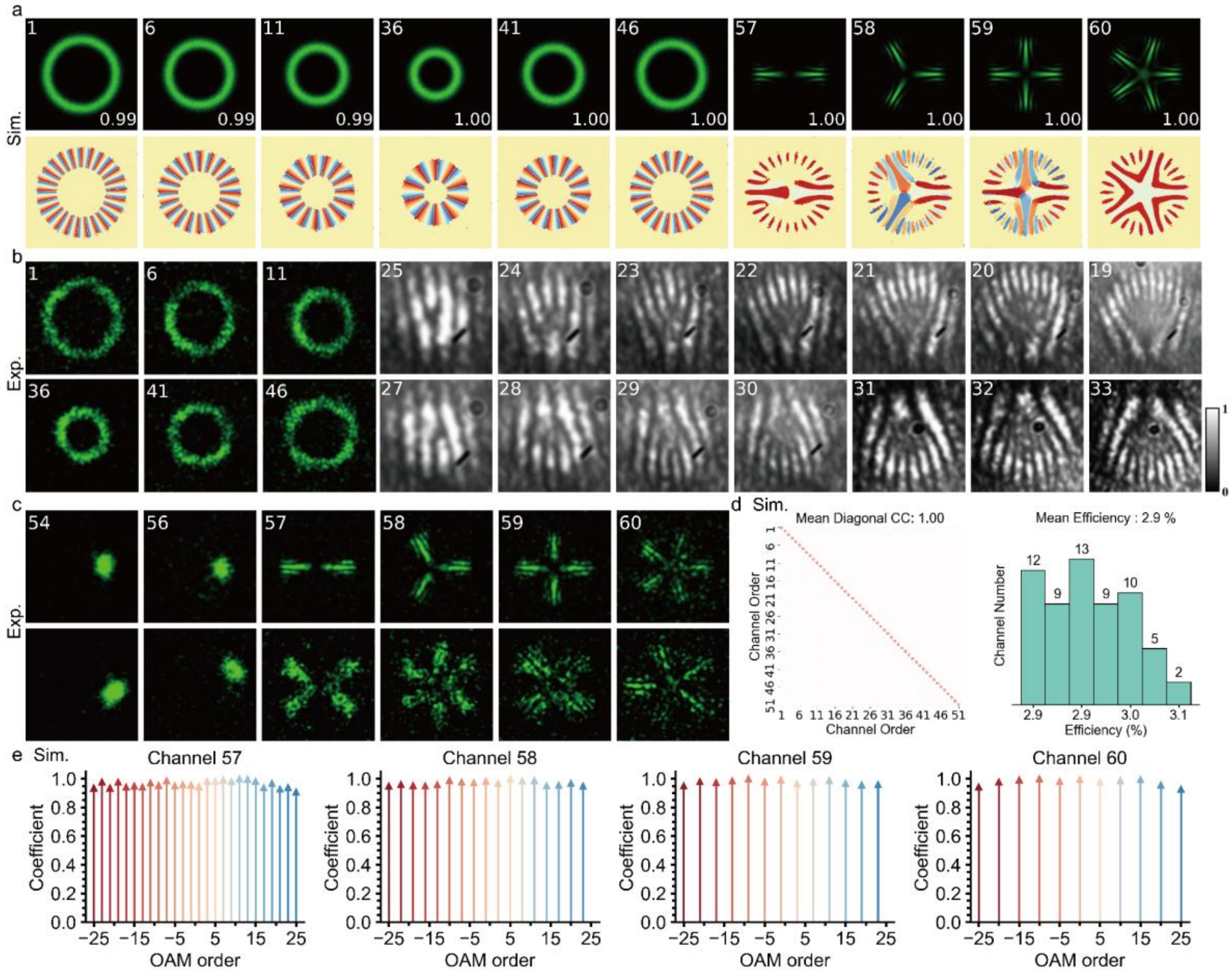

**Fig. 4 | Results of mechanically reconfigurable OAM generation. a** Simulated results of 10 representative channels from the 60-channel OAM beams. **b** Experimental results of 6-channel single-mode OAM intensities and 14-channel interfence pattern. **c** Experimental results of slected 6-channel multi-mode OAM intensities in target plane and 8-cm transimision plane. **d** CC matrix of generated 51-channel OAM beams and efficiency histogram of all channels. **e** Normalised OAM coefficient distributions of multi-mode OAM beams in channels 57, 58, 59, 60.

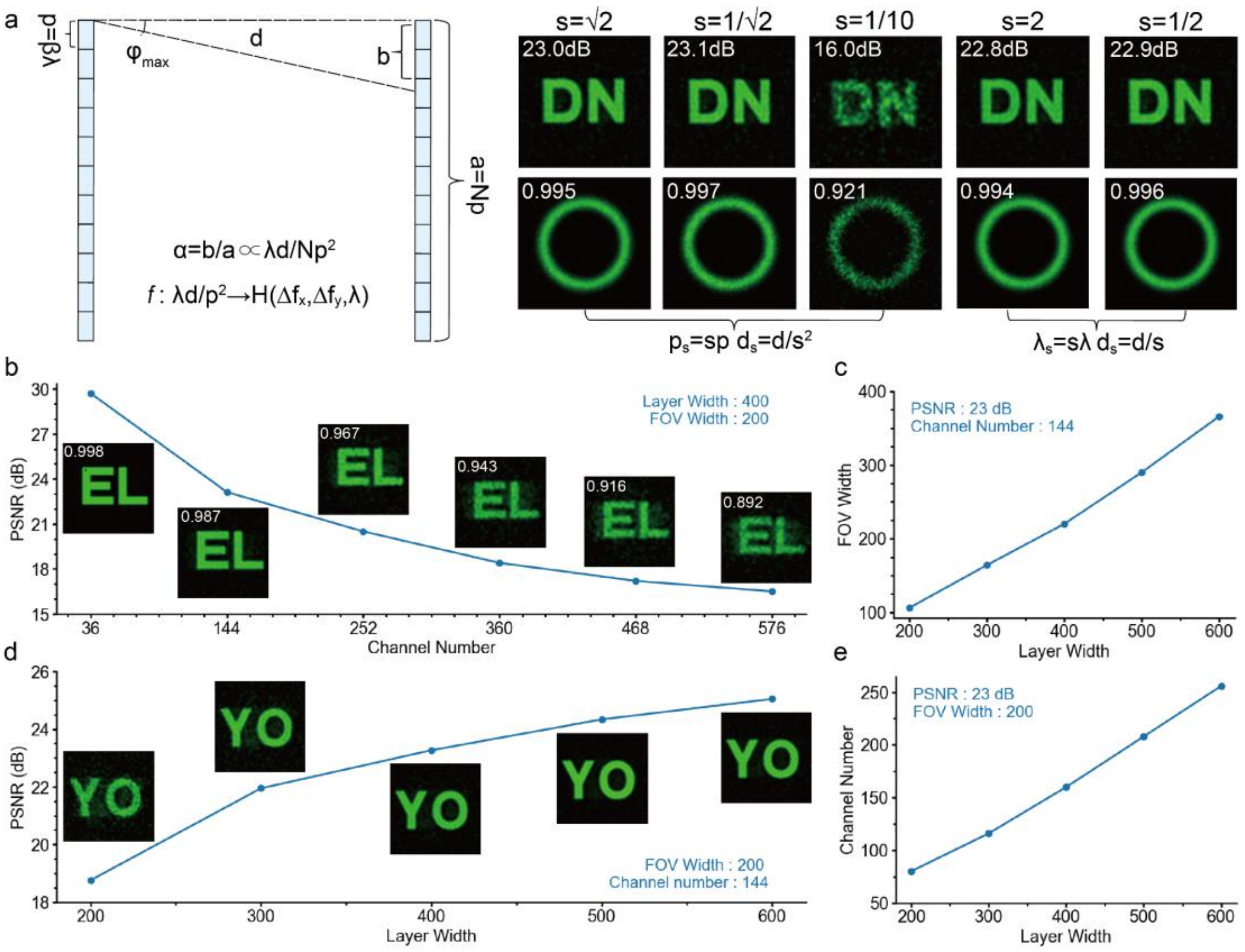

**Fig. 5 | The equivalent connection law and model scalability of DMCN. a** Side-view schematic of the diffraction connection between two adjacent modulation layers and the output optical fields of the identical connection models. The top-left corner of the image is labeled with the average PSNR or average CC. **b-e** Line charts analyzing the relationships among Layer Width, FOV Width, PSNR, and Channel Number, based on the selected 576 channels. The letter images are used as target holograms. We employ the controlled variable method, fixing the values of two variables while discussing the interactions between the other two.